\documentclass[conference]{IEEEtran}
\usepackage{graphicx}
\usepackage{subfigure}
\usepackage{float}
\usepackage{algorithm}
\usepackage{algorithmic}
\usepackage{mathrsfs}
\usepackage{amsfonts}
\usepackage{array}
\usepackage{fancyhdr}
\usepackage{amssymb,amsmath}
\usepackage{longtable}
\usepackage{rotating}
\usepackage{multirow}
\usepackage{epstopdf}
\usepackage{cite}
\usepackage{float}
\usepackage[marginal]{footmisc}
\usepackage{stfloats}
\usepackage[colorlinks,
linkcolor=black,
anchorcolor=black,
urlcolor=blue,
citecolor=black]{hyperref}
\usepackage[hyphenbreaks]{breakurl}

\renewcommand{\baselinestretch}{1}

\def\BibTeX{{\rm B\kern-.05em{\sc i\kern-.025em b}\kern-.08em
		T\kern-.1667em\lower.7ex\hbox{E}\kern-.125emX}}

\renewcommand{\baselinestretch}{1}
\newtheorem{definition}{Definition}

\begin{document}
	
	\title{A Stackelberg Game Approach to Resource Allocation for IRS-aided Communications}
	\author{\IEEEauthorblockN{Yulan Gao\textsuperscript{1,2}, Chao Yong\textsuperscript{1}, Zehui Xiong\textsuperscript{2}, Dusit Niyato\textsuperscript{2},  Yue Xiao\textsuperscript{1}, Jun Zhao\textsuperscript{2}}
		
		\IEEEauthorblockA{\textsuperscript{1}{The National Key Laboratory of Science
				and Technology on Communications} \\
			{University of Electronic
				Science and Technology of China, Chengdu 611731, China }\\
			{email: xiaoyue@uestc.edu.cn}}
		
		\IEEEauthorblockA{\textsuperscript{2}{The School of Computer Science and Engineering,}
			{Nanyang Technological University, Singapore  639798}}
	}

	\maketitle

 \thispagestyle{fancy}
\pagestyle{fancy}
\lhead{This paper appears in the Proceedings of IEEE Global Communications Conference (\textbf{GLOBECOM}) 2020.\\ Please feel free to contact us for questions or remarks. \hfill \url{http://junzhao.info/} }
\cfoot{\thepage}
\renewcommand{\headrulewidth}{0.4pt}
\renewcommand{\footrulewidth}{0pt}

\begin{abstract}
It is known that the capacity of the intelligent reflecting surface (IRS) aided cellular network can be effectively improved by reflecting the incident signals from the transmitter in a low-cost passive reflecting way.
Nevertheless, in the actual network operation, the base station (BS) and IRS may belong to different operators, consequently, the IRS is reluctant to help the BS without any payment.
Therefore, this paper investigates price-based reflection resource (elements) allocation strategies for an IRS-aided multiuser multiple-input and single-output (MISO) downlink communication systems, in which all transmissions over the same frequency band.
Assuming that the IRS is composed with multiple modules, each of which is attached with a smart controller, thus, the states (active/idle) of module can be operated by its controller, and all controllers can be communicated with each other via fiber links.
A Stackelberg game-based alternating direction method of multipliers (ADMM) is proposed to jointly optimize the transmit beamforming at the BS and the passive beamforming of the active modules.
Numerical examples are presented to verify the proposed algorithm.
It is shown that the proposed scheme is effective in the utilities of both the BS and IRS.
\end{abstract}

\IEEEpeerreviewmaketitle

\section{Introduction}
\IEEEPARstart{B}{y enabling} the intelligent reflecting surface (IRS) to the wireless systems, the IRS-aided wireless system recently has attracted significant interest due to its potential to further improve the system capacity and spectral efficiency \cite{Wu2019Beamforming, Zenzo2019Smart}.
Specifically, IRS exploits large reflecting elements to proactively steer the incident radio-frequency wave towards destination terminals \cite{Basar2019Wireless}, which is a promising solution to build a programmable wireless environment for 6G systems \cite{Hu2019Reconfigurable}.
Thereby, the fine-grained three-dimensional reflecting beamforming can be achieved without the need of any transmit radio frequency (RF) chain \cite{Han2019Large}.

\subsection{Related Work}
The IRS-aided wireless systems refer to the scenario that a large number of software-controlled reflection elements  with adjustable phase shifts for reflecting the incident signal.
As such,  the phase shifts of all reflecting elements can be tuned adaptively according to the state of networks, e.g., the channel conditions and the incident angle of the signal by the base station (BS).
It is commonly believed that the propagation environment can be improved without incurring additional noise at the reflector elements.
Currently, major communication field researchers are actively involved in the research of IRS-aided communications \cite{Zenzo2019Reconfigurable, Guo2019Weighted,Wu2019Intelligentjournal}.
For example,  \cite{Zenzo2019Reconfigurable} summarized the main communication applications and competitive advantages of the IRS technology.
In the spirit of these works, a vast corpus of literature focused on optimizing active-passive beamforming for unilateral spectral efficiency maximization subject to power constraint.
For instance, \cite{Guo2019Weighted} proposed a fractional programming based alternating optimization approach to maximize the weighted SE in IRS-aided multiple-input multiple-output (MISO) downlink communication systems.
In particular, three assumptions for the feasible set of reflection coefficient  were consider at IRS, including the ideal reflection coefficient constrained by peak-power, continuous phase shifter, and discrete phase shifter.
Meantime, in MISO wireless systems, the problem of minimizing the total transmit power at the access point was considered to
energy-efficient active-passive beamforming \cite{Wu2019Intelligentjournal}.
Notably, the aforementioned studies for IRS-aided communications were based on the premise of ignoring the power consumption at IRS.
In contrast, in \cite{Huang2019Reconfigurable},  an energy efficiency (EE) maximization problem was investigated by developing a realistic IRS power consumption model, where IRS power consumption relies on the type and the resolution of meta-element.

\subsection{ Motivation and Contributions}

The above resource allocation works address the joint transmit beamforming and phase shift optimization problem in IRS-aided communication systems.
These works assume that IRS operators are all selfless, and will always participate in the cooperative transmission despite their own energy consumption/maintanence cost \cite{Huang2019Reconfigurable} and profits.
However, this assumption becomes unrealistic in practice, due to the advances in intelligent communication and the shrinking resources.
In other words, if an IRS operator cannot benefit from the participation, it will not join in the cooperative communication.
Moreover, the  common assumption in the existing studies for IRS-aided communications is that all the reflecting elements are used to reflect the incident signal, i.e., adjusting reflecting coefficient of each meta-element simultaneously each time.
However, along with the use of a large number of high-resolution reflecting elements, especially with continuous phase shifters, triggering all the reflecting elements every time may result in significant  power consumption \cite{Huang2019Reconfigurable}.
Moreover, the hardware support for the  IRS implementation is the use of a large number of tunable metasurfaces.
Specifically, the tunability feature can be realized by introducing mixed-signal integrated circuits (ICs) or diodes/varactors, which can  vary both the resistance and reactance, offering complete
local control over the complex surface impedance \cite{Liu2019Intelligent}.
According to the IRS power consumption model presented in \cite{Huang2019Reconfigurable} and the hardware support, activating the entire IRS not only incurs increased power consumption, but also entails the increased latency of adjusting phase-shift and accelerates equipment depreciation.
Therefore, realizing  reflection resource management is significantly important for IRS-aided communications.
In this paper, for IRS-aided multiuser MISO systems, we consider the resource allocation problem in which an IRS operator serves the BS and prices the active modules.
The problem is formulated as a Stackelberg game, in which the IRS operator decides the price for the active modules.

The contributions of this paper are summarized as follows:
\begin{itemize}
\item{For the first time,  a modular architecture of IRS is proposed that  divides all the reflecting elements into multiple modules, whose states (active and idle) are controlled by multiple parallel switches belonging to a common controller.  We assume that each module contains multiple reflecting elements, i.e., the size of each module is larger than the incident signal wavelength, since the unit meta-element size is subwavelength \cite{Liu2019Intelligent}.
    As mentioned in \cite{Zenzo2019Smart},  the IRS is programmatically controlled by the controller, and hence,  from an operational standpoint, independent module activating can be implemented easily.
    Therefore, the proposed architecture of IRS allows the realization of the  reflection resource management, since each module is independently activated by its switch.}

\item{Based on the proposed modular architecture of IRS, this paper proposes a new price-based resource allocation scheme for both the BS and IRS.
    Furthermore, the Stackelberg game is formulated to maximize the individual revenue of the BS and IRS for the proposed price-based resource allocation.
    Since the entire game is a non-convex mixed-integer problem, which is even hard to solve in a centralized way, the problem is transformed into a convex problem by introducing the mixed row block $\ell_{1,2}\text{-norm}$ \cite{Mehanna2013Joint}, which yields a suitable semidefinite relaxation.
    To solve this problem, we apply a Stackelberg game-based alternating direction method of multipliers (ADMM) to identify the price, active module subsets, and subsequently both the transmit power allocation and the corresponding passive beamforming.
    }
\end{itemize}


\section{System Model and Problem Formulation}
\subsection{System Model}
Consider the downlink communication between a BS equipped with $M$ antennas and $K$ single-antenna mobile users. The communication takes place via an IRS with $S$ modules, and each module consisting $N$ reflection elements, and thus,  the total reflecting elements of IRS is $SN.$
Define ${\cal K}:=\{1, 2, \ldots, K\},$ ${\cal S}:=\{1, 2, \ldots, S\},$ and ${\cal I}=\{1, 2, \ldots, (SN)\}$ as the index sets of users, the reflection modules, and the reflecting elements, respectively.
Let ${\mathbf H}_{0,s}\in{\mathbb C}^{N\times M}$ be the channel matrix from the BS to the $s\text{th}$  module of IRS, ${\mathbf g}_{s,k}\in{\mathbb C}^{N\times 1}$ be the channel vector from the $s\text{th}$  module of the IRS to user $k.$
The direct channel for the BS to user $k$ is denoted as $h_{d,k}\in{\mathbb C}^{M\times 1}.$
Denote by $\phi_i, \forall i\in{\cal I}$ the $i\text{th}$ reflecting element of the IRS.
Let ${\pmb\Phi}=\text{diag}\{{\pmb\Phi}_1, {\pmb\Phi}_2, \ldots, {\pmb\Phi}_S\}\in{\mathbb C}^{(SN)\times(SN)},$ where ${\pmb\Phi}_s=\text{diag}[\phi_{(s-1)N+1}, \phi_{(s-1)N+2}, \ldots, \phi_{sN}]\in{\mathbb C}^{N\times N}.$
Define ${\pmb\phi}=[({\pmb\phi}_1)^T,({\pmb\phi}_2)^T,\ldots, ({\pmb\phi}_S)^T ]^T\in{\mathbb C}^{(SN)\times 1}, $  where ${\pmb\phi}_s=[({\phi}_{(s-1)N+1})^{\dag}, ({\phi}_{(s-1)N+2})^{\dag}, \ldots, ({\phi}_{sN})^{\dag}]^T\in{\mathbb C}^{N\times 1}.$

We assume that all the  modules of IRS can potentially join the cooperative communication, then, the channel matrix from the BS to the IRS and the IRS to user $k$ respectively are
\begin{equation}\label{eq:1}
\begin{aligned}
{\mathbf H}&=[({\mathbf H}_{0,1})^T, ({\mathbf H}_{0,2})^T, \ldots, (  {\mathbf H}_{0,S})^T   ]^T\in{\mathbb C}^{(SN)\times M}\\
{\mathbf g}_k&=[ ({\mathbf g}_{1,k})^T, ({\mathbf g}_{2,k})^T, \ldots, ({\mathbf g}_{S,k})^T ]^T\in{\mathbb C}^{(SN)\times 1}, \forall k\in{\cal K}.
\end{aligned}
\end{equation}
The SINR for user $k$, which is denoted by $\gamma_k$ can be computed by
\begin{equation}\label{eq:2}
\gamma_k=\frac{|({\mathbf h}_{d,k}^{\dag}+{\mathbf g}_{k}^{\dag}{\pmb\Phi}(||\pmb\Phi||_{0,F}){\mathbf H}){\mathbf w}_k |^2}
{\sum_{j\neq k}^K |({\mathbf h}_{d,k}^{\dag}+{\mathbf g}_k^{\dag}{\pmb\Phi}(||{\pmb\Phi}||_{0,F}){\mathbf H}){\mathbf w}_j |^2+\sigma^2},
\end{equation}
where ${\mathbf w}_k\in{\mathbb C}^{M\times 1}$ is the transmit beamforming vector for user $k,$
$\sigma^2$ is the background noise at user $k$, and the $\ell_{0,F}\text{-norm}$ is the number of nonzero blocks of matrix $\pmb\Phi,$ i.e., $||{\pmb\Phi}||_{0,F}\triangleq \left|\left\{s: ||{\pmb\Phi}_s||_F\neq 0  \right\}\right|.$
Moreover, $\pmb\Phi(||\pmb\Phi||_{0,F})$ is the corresponding phase shift for the active modules.


\subsection{Stackelberg Game Formulation}
In this paper, we assume that the BS and IRS belong to different operators, and the IRS is selfish.
In the case of unfavorable propagation condition on the direct signal path,  in order to improve the sum rate of system, the BS needs to pay for the IRS's forwarding service.
Then, the IRS's objective is to maximize its utility, denoted by $V$, calculated as follows:
\begin{equation}\label{eq:7}
V={ r}||\pmb\Phi||_{0,F},
\end{equation}
where $r$ is the price to IRS for providing $||\pmb\Phi||_{0,F}$ reflection modules.
Notably, in IRS-aided communication, the authority to adjust the passive beamforming, $\pmb\Phi(||\pmb\Phi||_{0,F})$ is controlled by the BS.
The objective of the IRS is to solve the following problem:
\begin{equation}\label{s:1}
\text{Leader-Problem:}~~\max_{r}~V,~~~\text{s.t.~~} r>0.
\end{equation}

In response to the action of the IRS, the BS chooses a best $||\pmb\Phi||_{0,F}$ active modules, decides the phase shift $\pmb\Phi(||\pmb\Phi||_{0,F})$ of the activated modules and its own transmit beamforming ${\mathbf W}.$ Then, the BS's utility is designed as the sum data rate of all users excluding its cost of forwarding service, which is formulated as
\begin{equation}\label{eq:3}
\begin{aligned}
U=\sum\nolimits_{k=1}^K \log_2\left( 1+\gamma_k\right)-{r}||{\pmb\Phi}||_{0,F}.
\end{aligned}
\end{equation}
The problem of obtaining the optimal strategy for the BS can be formulated as follows:
\begin{subequations}\label{s:2}
\begin{align}
\text{Follower-Problem:}~~&\max_{{\mathbf w}_k, {\pmb\Phi}}~~~U\\
\text{s.t.~}& \sum\nolimits_{k=1}^K ||{\mathbf w}_k||_2^2\leq p^{\max}\\
&|\phi_i|\leq 1, \forall i=1,2, \ldots, (SN).
\end{align}
\end{subequations}

Since optimization problems (\ref{s:1}) and (\ref{s:2}) is nonconvex and generally impossible to solve as their solutions usually requires an intractable combinational search.
A common alternative is to consider the mixed $\ell_{1,2}\text{-norm}$, defined as
$||\pmb\Phi||_{1,F}=\sum_{s=1}^S||{\pmb\Phi}_{s}||_F.$
Note that the $\ell_{1,F}\text{-norm}$ behaves as the $\ell_{0,F}\text{-norm}$ on $\pmb\Phi$, which implies that each $||{\pmb\Phi}_s||_F$ is encouraged to be zero, therefore inducing group-sparsity \cite{CandEnhancing2007}.
For our purpose, we will use the convex $\ell_{1,F}\text{-norm}$ as a group-sparsity to replace the nonconvex $\ell_{0,F}\text{-norm}$ in (\ref{s:1}) and (\ref{s:2}).
Through all the transformation, the objectives of the leader and follower can thus be relaxed to
\begin{subequations}\label{s:3}
\begin{align}
\text{F-Problem:}~&\max_{{\mathbf w}_k, \pmb\Phi}\sum\nolimits_{k=1}^K\text{log}_2(1+\gamma_k)-r\delta||\pmb\Phi||_{1,F}\\
\text{s.t.}~~&(\ref{s:2}\text{b})  ~\text{and}~(\ref{s:2}\text{c}),
\end{align}
\end{subequations}
where $\delta$ is the positive real tuning parameter that controls the sparsity of the solution, and thus the number of activated modules.
\begin{equation}\label{s:4}
\text{L-Problem:}~\max_{r}~r\delta||\pmb\Phi||_{1,2}, ~\text{s.t.~} r>0.
\end{equation}

For the proposed Stackelberg game, the Stackelberg game equilibrium (SE) is defined as follows.
\begin{definition}
Define $ {\mathbf W}=[{\mathbf w}_1, {\mathbf w}_2, \ldots,{\mathbf w}_K]\in{\mathbb C}^{M\times K}.$
Let ${ r}^{*}$ be a solution of problem (\ref{s:4}) and $({\mathbf W}^{*}, {\pmb\Phi}^{*})$ be a solution for problem (\ref{s:3}). Then, the point $({ r}^{*}, {\mathbf W}^{*}, {\pmb\Phi}^{*})$ is the Stackelberg equilibrium for the proposed Stackelberg game if for any $({ r}, {\mathbf W}, {\pmb\Phi})$, the following conditions are satisfied:
\begin{equation}\label{eq:10}
\begin{aligned}
U({ r}^{*}, {\mathbf W}^{*}, {\pmb\Phi}^{*})&\geq U({ r}^{*}, {\mathbf W}, {\pmb\Phi})\\
V({ r}^{*}, {\mathbf W}^{*}, {\pmb\Phi}^{*})&\geq V({ r}, {\mathbf W}^{*}, {\pmb\Phi}^{*}).
\end{aligned}
\end{equation}
\end{definition}

\section{Game Analysis}
In the proposed game, both at the BS's and the IRS's side, since there is only one player, the best response of the BS and IRS can be readily obtained by solving { F-Problem} and { L-Problem}, respectively.
For the proposed game, the SE can be obtained as follows: For a given ${ r},$ { F-Problem } (\ref{s:3}) is solved first. Then, with the obtained best response functions $({ W}^{*}, {\pmb\Phi}^{*})$ of the BS, we solve {L-Problem} (\ref{s:4}) for the optimal price ${ r}^{*}.$

\subsection{Strategy Analysis for the BS}

If we denote the price for serving the BS as ${ r},$ the optimization problem (\ref{s:3}) can be solved by treating parameter $r$ as constant.

To tackle the logarithm in the objective function of (\ref{s:3}), we apply the Lagrangian dual transform.
Then, the objective function of (\ref{s:3}) can be equivalently written as
\begin{equation}\label{eq:12}
\begin{aligned}
\max_{{\mathbf W}, {\pmb\Phi}}~\sum\nolimits_{k=1}^K\log_2\left(1+\alpha_k\right)&-\sum\nolimits_{k=1}^K\alpha_k
+\sum\nolimits_{k=1}^K\frac{(1+\alpha_k)\gamma_k}{1+\gamma_k}\\
&-{ r}\delta\sum\nolimits_{s=1}^S||{\pmb\Phi}_s||_F.
\end{aligned}
\end{equation}
In (\ref{eq:12}), when ${\mathbf W}$ and ${\pmb\Phi}$ hold fixed, the optimal $\alpha_k$ is
$\alpha_k^{*}=\gamma_k, \forall k\in{\cal K}.$
Then, for a given price ${ r}$ and  a fixed $\{\alpha_k\}_{k\in{\cal K}},$ optimizing ${\mathbf W}$ and ${\pmb\Phi}$ is reduced to
\begin{equation}\label{eq:14}
\begin{aligned}
\max_{{\mathbf W}, {\pmb\Phi}} \sum\nolimits_{k=1}^K\frac{\tilde{\alpha}_k\gamma_k}{1+\gamma_k}-{ r}\delta\sum\nolimits_{s=1}^S||{\pmb\Phi}_s||_F,
\end{aligned}
\end{equation}
where $\tilde{\alpha}_k=1+\alpha_k.$

\subsubsection{Transmit Beamforming}
In the following, we investigate how to find a better beamforming matrix ${\mathbf W}$ given fixed ${\pmb\Phi}$ for (\ref{eq:14}). Denote the combined channel for user $k$ by
\begin{equation}\label{eq:15}
{\mathbf h}_k^{\dag}={\mathbf h}_{d,k}^{\dag}+{\mathbf g}_k^{\dag}{\pmb\Phi}{\mathbf H}, \forall k\in{\cal K.}
\end{equation}
Then, the SINR $\gamma_k$ in (\ref{eq:2}) is given by
\begin{equation}\label{eq:16}
\gamma_k=\frac{|{\mathbf h}_k^{\dag}{\mathbf w}_k|^2}
{\sum\nolimits_{j\neq k}^K|{\mathbf h}_k{\mathbf w}_j|^2+\sigma^2}.
\end{equation}
Using $\gamma_k$ in (\ref{eq:16}), the objective function of (\ref{eq:14}) is written as a function of $\{{\mathbf w}_k\}_{k=1}^K:$
\begin{equation}\label{eq:17}
\begin{aligned}
\sum_{k=1}^K\frac{\tilde{\alpha}_k\gamma_k}{1+\gamma_k}-&{ r}\delta\sum_{s=1}^S||{\pmb\Phi}_s||_F
=\\&\sum_{k=1}^K\frac{\tilde{\alpha}_k|{\mathbf h}_k^{\dag}{\mathbf w}_k|^2}
{\sum_{j=1}^K|{\mathbf h}_k^{\dag}{\mathbf w}_j|^2+\sigma^2}-{ r}\delta\sum_{s=1}^S||{\pmb\Phi}_s||_F.
\end{aligned}
\end{equation}
Thus, for given ${ r},$ $\{{\alpha}_k\}_{k\in{\cal K}},$ and ${\pmb\Phi}$, optimizing $\{{\mathbf w}_k\}_{k=1}^K$ becomes
\begin{equation}\label{eq:18}
\begin{aligned}
\max_{\{{\mathbf w}_k\}_{k=1}^K}& \sum_{k=1}^K\frac{\tilde{\alpha}_k|{\mathbf h}_k^{\dag}{\mathbf w}_k|^2}
{\sum_{j=1}^K|{\mathbf h}_k^{\dag}{\mathbf w}_j|^2+\sigma^2}\\
\text{s.t.~}&\sum_{k=1}^K||{\mathbf w}_k||_2^2\leq p^{\max}.
\end{aligned}
\end{equation}
Using quadratic transform, the objective function of (\ref{eq:18}) is reformulated as
\begin{equation}\label{eq:19}
\begin{aligned}
\sum_{k=1}^K&\frac{\tilde{\alpha}_k|{\mathbf h}_k^{\dag}{\mathbf w}_k|^2}
{\sum_{j=1}^K|{\mathbf h}_k^{\dag}{\mathbf w}_j|^2+\sigma^2}
=\sum\nolimits_{k=1}^K2\sqrt{\tilde{\alpha}_k}\text{Re}\left\{ \beta_k^{\ddag}{\mathbf h}_k^{\dag}{\mathbf w}_k \right\}\\
&-\sum\nolimits_{k=1}^K|\beta_k|^2\left(\sum\nolimits_{j=1}^K|{\mathbf h}_k^{\dag}{\mathbf w}_j|^2+\sigma^2\right),
\end{aligned}
\end{equation}
where $(\cdot)^{\ddag}$ denotes the conjugate. $\beta_k\in{\mathbb C}$ is the auxiliary variable. Then, solving problem (\ref{eq:18}) over $\{{\mathbf w}_k\}_{k=1}^K$ is equivalent to solving the following problem over $\{{\mathbf w}_k\}_{k=1}^K$ and ${\pmb\beta}=[\beta_1, \ldots, \beta_K]^T\in{\mathbb C}^{K\times 1}:$
\begin{equation}\label{eq:20}
\begin{aligned}
\max_{\{{\mathbf w}_k\}_{k=1}^K, {\pmb\beta}}~&~\sum\nolimits_{k=1}^K2\sqrt{\tilde{\alpha}_k}\text{Re}\left\{ \beta_k^{\ddag}{\mathbf h}_k^{\dag}{\mathbf w}_k \right\}\\
&-\sum\nolimits_{k=1}^K|\beta_k|^2\left(\sum\nolimits_{j=1}^K|{\mathbf h}_k^{\dag}{\mathbf w}_j|^2+\sigma^2  \right)\\
\text{s.t.~}&\sum\nolimits_{k=1}^K||{\mathbf w}_k||_2^2\leq p^{\max}.
\end{aligned}
\end{equation}
The optimal $\beta_k$ for a given $\{{\mathbf w}_k\}_{k=1}^K$ is
\begin{equation}\label{eq:21}
\beta_k^{*}=\frac{\sqrt{\tilde{\alpha}_k}{\mathbf h}_k^{\dag}{\mathbf w}_k}
{\sum_{j=1}^K|{\mathbf h}_k^{\dag}{\mathbf w}_j|^2+\sigma^2}.
\end{equation}
Then, fixing ${\pmb\beta},$ the optimal ${\mathbf w}_k$ is
\begin{equation}\label{eq:22}
{\mathbf w}_k^{*}=\sqrt{\tilde{\alpha}_k}\beta_k(\lambda_0{\mathbf I}_M+\sum\nolimits_{j=1}^K|\beta_j|^2{\mathbf h}_j{\mathbf h}_j^{\dag})^{-1}{\mathbf h}_k,
\end{equation}
where $\lambda_0$ is the dual variable introduced for the power constraint, which is optimally determined by
\begin{equation}\label{eq:23}
\lambda_0^{*}=\max\{0, p^{\max}-\sum\nolimits_{k=1}^K||{\mathbf w}_k||_2^2 \}.
\end{equation}
\subsubsection{Optimizing Reflection Response Matrix ${\pmb\Phi}$}
Optimize ${\pmb\Phi}$ in (\ref{eq:14}) given fixed pricing ${\mathbf r},$ $\{\alpha_k\}_{k\in{\cal K}},$ and $\{{\mathbf w}_k\}_{k=1}^K$.
Using $\gamma_k$ defined in (\ref{eq:2}), the objective function of (\ref{eq:14}) is expressed as a function of ${\pmb\Phi}$:
\begin{equation}\label{eq:24}
\sum_{k=1}^K\frac{\tilde{\alpha}_k|({\mathbf h}_{d,k}^{\dag}+{\mathbf g}_k^{\dag}{\pmb\Phi}(\mathbf H)){\mathbf w}_k|^2}
{\sum_{j=1}^K|({\mathbf h}_{d,k}^{\dag}+{\mathbf g}_k^{\dag}{\pmb\Phi}{\mathbf H}){\mathbf w}_j|^2+\sigma^2}-{ r}\delta\sum_{s=1}^S||{\pmb\Phi}_s||_F.
\end{equation}
Define ${\mathbf a}_{j,k}=\text{diag}\{{\mathbf g}_k^{\dag}\}{\mathbf H}{\mathbf w}_j, b_{j,k}={\mathbf h}_{d,k}^{\dag}{\mathbf w}_j,
\forall k,j=1, 2, \ldots, K.$
Combining with the definition of ${\pmb\phi}$, (\ref{eq:24}) can be rewritten as
\begin{equation}\label{eq:25}
\sum_{k=1}^K\frac{\tilde{\alpha}_k|b_{k,k}+{\pmb\phi}^{\dag}{\mathbf a}_{k,k}|^2}
{\sum_{j=1}^K|b_{j,k}+{\pmb\phi}^{\dag}{\mathbf a}_{j,k}|^2+\sigma^2}-{ r}\delta\sum_{s=1}^S||{\pmb\Phi}_s||_F.
\end{equation}
Note that $\sum_{s=1}^S||{\pmb\Phi}_s||_F=\sum_{s=1}^S||{\pmb\phi}_s||_2,$ optimizing ${\pmb\phi}$ can be represented as follows:
\begin{equation}\label{eq:26}
\begin{aligned}
\max_{{\pmb\phi}}~&\sum_{k=1}^K\frac{\tilde{\alpha}_k|b_{k,k}+{\pmb\phi}^{\dag}{\mathbf a}_{k,k}|^2} {\sum_{j=1}^K|b_{j,k}+{\pmb\phi}^{\dag}{\mathbf a}_{j,k}|^2+\sigma^2}-{ r}\delta\sum_{s=1}^S||{\pmb\phi}_s||_2\\
\text{s.t.~} &{\pmb\phi}^{\dag}{\mathbf e}_i{\mathbf e}_i^{\dag}{\pmb\phi}\leq 1, \forall i=1, 2, \ldots, (SN).
\end{aligned}
\end{equation}
Based on the quadratic transform, the new objective function of (\ref{eq:26}) is
\begin{equation}\label{eq:27}
\begin{aligned}
&\sum\nolimits_{k=1}^K2\sqrt{\tilde{\alpha}_k}\text{Re}\left\{ \epsilon_k^{\ddag}{\pmb\phi}^{\dag}{\mathbf a}_{k,k} +\epsilon_k^{\ddag}b_{k,k}\right\}-\sum\nolimits_{k=1}^K|\epsilon_k|^2\\
&\times\left( \sum\nolimits_{j=1}^K|b_{j,k}+{\pmb\phi}^{\dag}{\mathbf a}_{j,k}|^2+\sigma^2  \right)-{ r}\delta\sum\nolimits_{s=1}^S||{\pmb\phi}_s||_2^2,
\end{aligned}
\end{equation}
and ${\pmb\epsilon}=[\epsilon_1, \ldots, \epsilon_K]^T\in{\mathbb C}^{K\times 1}$ refers to the auxiliary variable vector.
Similarly, we optimize ${\pmb\phi}$ and $\pmb\epsilon$ alternatively \cite{Guo2019Weighted}.
The optimal $\epsilon_k$ for given $\pmb\phi$ can be obtained easily, shown as follows:
\begin{equation}\label{eq:28}
\epsilon_k^{*}=\frac{\sqrt{\tilde{\alpha}_k}(b_{k,k}+{\pmb\phi}^{\dag}{\mathbf a}_{k,k})}
{\sum_{j=1}^K|b_{j,k}+{\pmb\phi}^{\dag}{\mathbf a}_{j,k}|^2+\sigma^2}.
\end{equation}
Then, the remaining problem is optimizing $\pmb\phi$ for given $\pmb\epsilon.$
By introducing new variable ${\pmb\theta}={\pmb\phi}\in{\mathbb C}^{(SN)\times 1}.$
Likewise, ${\pmb\theta}_s\in{\mathbb C}^{N\times 1}$ represents the $s$th block of vector ${\pmb\theta}.$
Thus, for the fixed ${\pmb\epsilon},$ the optimization problem of ${\pmb\phi}$ is given as follows:
\begin{equation}\label{eq:29}
\begin{aligned}
\max_{{\pmb\phi}, {\pmb\theta}}~&\sum_{k=1}^K2\sqrt{\tilde{\alpha}_k}\text{Re}\left\{ \epsilon_k^{\ddag}{\pmb\phi}^{\dag}{\mathbf a}_{k,k}+\epsilon_k^{\ddag}b_{k,k}\right\}-\sum_{k=1}^K|\epsilon_k|^2\\
&\left( \sum_{j=1}^K|b_{j,k}+{\pmb\phi}^{\dag}{\mathbf a}_{j,k}|^2+\sigma^2  \right)-{ r}\delta\sum_{s=1}^S||{\pmb\theta}_s||_2^2\\
\text{s.t.}~& {\pmb\phi}^{\dag}{\mathbf e}_i{\mathbf e}_i^{\dag}{\pmb\phi}\leq 1, \forall i=1, 2, \ldots, (SN)\\
&{\pmb\theta}={\pmb\phi}.
\end{aligned}
\end{equation}

Utilizing the method of augmented Lagrangian minimization, (\ref{eq:29}) can be handled by solving
\begin{equation}\label{eq:30}
\begin{aligned}
\min_{{\pmb\Lambda}}\max_{{\pmb\phi}, {\pmb\theta}}~&~ L_c({\pmb\phi}, {\pmb\theta}, {\pmb\Lambda})\\
\text{s.t.~}&~{\pmb\phi}^{\dag}{\mathbf e}_i{\mathbf e}_i^{\dag}{\pmb\phi}\leq 1, \forall i=1, 2, \ldots, (SN),
\end{aligned}
\end{equation}
where $c>0$ is the penalty factor; ${\pmb\Lambda}\in{\mathbb C}^{(SN)\times 1}$ is the Lagrangian vector multiplier for ${\pmb\theta}={\pmb\phi}.$
The partial augmented Lagrangian function is defined as
\begin{equation}\label{eq:31}
\begin{aligned}
L_c&({\pmb\phi},{\pmb\theta},{\pmb\Lambda})=\sum_{k=1}^K2\sqrt{\tilde{\alpha}_k}\text{Re}\left\{ \epsilon_k^{\ddag}{\pmb\phi}^{\dag}{\mathbf a}_{k,k}+\epsilon_k^{\ddag}b_{k,k} \right\}\\
&-\sum_{k=1}^K|\epsilon_k|^2\left( \sum_{j=1}^K|b_{j,k}+{\pmb\phi}^{\dag}{\mathbf a}_{j,k}|^2+\sigma^2  \right)\\
&-{ r}\delta\sum_{s=1}^S||{\pmb\theta}_s||_2^2-
\text{Re}\left\{\text{Tr}\left[{\pmb\Lambda}^{\dag}({\pmb\theta}-{\pmb\phi}) \right]\right\}
-\frac{c}{2}||{\pmb\theta}-{\pmb\phi}||_2^2.
\end{aligned}
\end{equation}
\begin{itemize}
\item{\textit{Updating} ${\pmb\phi}$:

By dual theory and KKT conditions, the optimal solution is given by (\ref{eq:32}).
{\small
\begin{figure*}
\begin{equation}\label{eq:32}
\begin{aligned}
{\pmb\phi}^{*}=\left(2\sum_{k=1}^K|\epsilon_k|^2 \sum_{j=1}^K {\mathbf a}_{j,k}{\mathbf a}_{j,k}^{\dag}+2\sum_{i=1}^{SN}\mu_i{\mathbf e}_i{\mathbf e}_i^{\dag}+c{\mathbf I}_{SN} \right)^{-1}
\left(2\sum_{k=1}^K\sqrt{\tilde{\alpha}_k}\epsilon_k^{\ddag}{\mathbf a}_{k,k}+{\pmb\Lambda}+c{\pmb\theta}-2\sum_{k=1}^K|\epsilon_k|^2\sum_{j=1}^Kb_{j,k}{\mathbf a}_{j,k}\right),\\
\hline
\end{aligned}
\end{equation}
\end{figure*}}
The Lagrangian multiplier $\mu_i$ updated by
\begin{equation}\label{eq:33}
\mu_i^{*}=\max\left\{0, 1-{\pmb\phi}^{\dag}{\mathbf e}_i{\mathbf e}_i^{\dag}{\pmb\phi}\right\}.
\end{equation}
}
\item{\textit{Updating } ${\pmb\theta}$:

The problem of ${\pmb\theta}$ is an unconstrained group  leastabsolute selection and shrinkage operator (group Lasso) problem \cite{Yuan2006Model}, i.e.,
\setcounter{equation}{31}
\begin{equation}\label{eq:34}
\max_{\pmb\theta}~-{r}\delta\sum_{s=1}^S||{\pmb\theta}_s||_2-\text{Re}\left\{\text{Tr}
\left[{\pmb\Lambda}^{\dag}({\pmb\theta}-{\pmb\phi})\right] \right\}-\frac{c}{2}||{\pmb\theta}-{\pmb\phi}||_2^2.
\end{equation}
Let ${\pmb\Lambda}_s\in{\mathbb C}^{N\times 1}$ denote the $s\text{th}$ row block of vector ${\pmb\theta}, s=1, 2, \ldots, S.$
Then, (\ref{eq:34}) can be divided into $S$ independent problems of ${\pmb\theta}_s$ for $s=1, 2, \ldots, S$
\begin{equation}\label{eq:35}
\max_{{\pmb\theta}_s}~-{ r}\delta||{\pmb\theta}_s||_2-\text{Re}\left\{\text{Tr}\left[{\pmb\Lambda}_s^{\dag}
({\pmb\theta}_s-{\pmb\phi}_s)  \right] \right\}-\frac{c}{2}||{\pmb\theta}_s-{\pmb\phi}_s||_2^2
\end{equation}
Defining ${\mathbf x}_s=c{\pmb\phi}_s-{\pmb\Lambda}_s,$ and ${\mathbf x}_s-c{\pmb\theta}_s\in{ r}\partial ||{\pmb\theta}_s||_2$, and thus,  we can easily obtain
\begin{equation}\label{eq:36}
{\pmb\theta}_s=\left\{ \begin{array}{ccc}
&{\mathbf 0}, & \text{~if~} ||{\mathbf x}_s||_2\leq { r\delta}\\
&\frac{(||{\mathbf x}_s||_2-{ r}\delta){\mathbf x}_s}
{c||{\mathbf x}_s||_2}, &\text{otherwise}.
\end{array}\right.
\end{equation}
The update of Lagrangian vector $\pmb\Lambda_s$ is given by
\begin{equation}\label{eq:37}
{\pmb\Lambda}_s={\pmb\Lambda}_s+c({\pmb\theta}_s-{\pmb\phi}_s), \forall s=1, 2, \ldots, S.
\end{equation}
}
\end{itemize}

\subsection{Game Analysis for the IRS Pricing}
Substituting (\ref{eq:36}) into ({ L-Problem}) in (\ref{s:4}), the optimization problem at the IRS side can be formulated as
\begin{equation}\label{eq:38}
\max_{{ r}>0}~ \sum_{s=1}^S\kappa_s\frac{-\delta^2{ r}^2+\delta||{\mathbf x}_s||_2{ r}}
{c},
\end{equation}
where $\kappa_s$ is indicate function, i.e.,
\begin{equation}\label{eq:39}
\kappa_s=\left\{\begin{array}{ccc}
&0, &\text{~if~} ||{\mathbf x}_s||_2\leq { r\delta}\\
&1, &\text{~otherwise}.
\end{array}\right.
\end{equation}
The optimal solution of (\ref{eq:38}) is
\begin{equation}\label{eq:40}
{ r}^{*}=\frac{\sum_{s=1}^S\kappa_s||{\mathbf x}_s||_2}
{2\delta\sum_{s=1}^S\kappa_s}.
\end{equation}

The entire framework including the identifying the price and the trigger module subsets as well as the transmit beamforming and the phase shift is summarized in Algorithm 1.
\begin{algorithm}[!htp]
\caption{ Algorithm Summary}             
\label{alg:Framwork}
\hspace*{0.02in} {\bf Initial:}  The IRS initialize the price ${ r}(1),$ and set the outer iteration number $\tau=1$   \\
\hspace*{0.02in} {\bf Part I:} the alternating optimization for solving (\ref{eq:14}) \\
(1.1) Initialize ${\mathbf W}(1)$ and ${\pmb\Phi}(1)$ to feasible values, and set the iteration number $t=1.$\\
\hspace*{0.02in} {\bf Repeat}\\
(1.2) Update the nominal SINR $\alpha_k(t), \forall k\in{\cal K}$;\\
(1.3) Update $\beta_k(t), \forall k\in{\cal K}$ by (\ref{eq:21});\\
(1.4) Update transmit beamforming ${\mathbf W}(t)$ by (\ref{eq:22}); update $\lambda_0(t)$ by (\ref{eq:23})\\
(1.5) Update $\epsilon_k(t), \forall k\in{\cal K}$ by (\ref{eq:28});\\
(1.6) Update ${\pmb\phi}(t)$ by (\ref{eq:32}); update $\mu_i(t), \forall i=1, 2, \ldots, (SN),$ by (\ref{eq:33});\\
(1.7) Update ${\pmb\theta}_s(t) $ by (\ref{eq:36}) in parallel for $s=1, 2,\ldots S;$\\
(1.8) Update $\pmb\Lambda_s(t)$ by (\ref{eq:37}) in parallel for $s=1, 2,\ldots, S;$\\
(1.9) Update $t=t+1;$\\
(1.10) {\bf Until } The value of function (\ref{eq:12}) converges.\\
\hspace*{0.02in} {\bf Part II:} Update price ${ r}$ by solving problem (\ref{eq:38}) in the outer loop \\
(2.1) Solve problem (\ref{eq:38}) for given $\{{\pmb\theta}_s(t)\}_{s=1}^S, \{{\pmb\Lambda}_s(t)\}_{s=1}^S, $  update ${ r}(\tau)$ by (\ref{eq:40})\\
(2.2) {\bf Until } the utility of the IRS is convergence.
\end{algorithm}

\section{Simulation Results}

In this section, extensive numerical results are presented to evaluate the performances of the proposed resource allocation strategies based on the approach of active module pricing. For simplicity, we set the balance parameter $\delta$ to $0.1.$
To keep the complexity of simulations tractable, we focus on the scenario, where the $K\in\{4, 6\}$ users are randomly deployed within a circle cell centered at $(200,0)\text{~m}$, and the cell radius is $10\text{~m}$, the BS and IRS are employed at $(0,0)\text{~m} $ and $(50, 50)\text{~m},$ respectively,  where the number of reflecting elements of each module is set as $N=8.$
We assume that the BS is equipped with $4$ ($6$) antennas for $K=4$ ($K=6$).
From \cite{Zheng2020Intelligent}, we set the path loss exponent of the direct link as $3.5$, and the path loss at the reference distance $1\text{~m}$ is set as $30\text{~dBm}$ for each individual link. For the IRS-aided link, $2$ is the value of the path loss exponent from the BS to the IRS and that from the IRS to users.
For simplicity, we assume the Rayleigh fading model to account for small-scale fading.
\begin{figure}[!t]
	\centering
	\begin{minipage}[t]{0.45\linewidth}
		\includegraphics[width=1.1\linewidth]{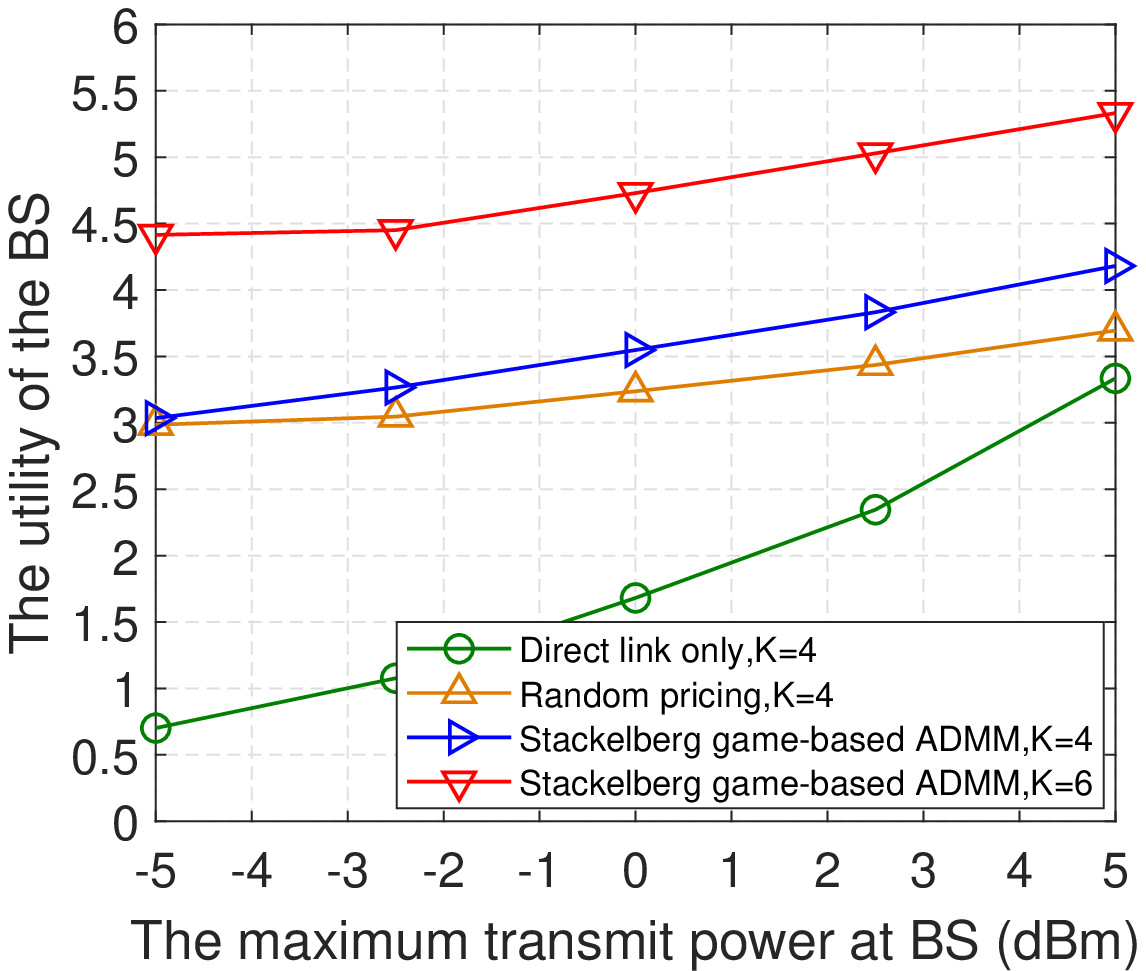}
		\caption{Impact of  $p^{\max}$ on $U$.}
		\label{fig:1}
	\end{minipage}
	\centering
	\begin{minipage}[t]{0.45\linewidth}
		\includegraphics[width=1.1\linewidth]{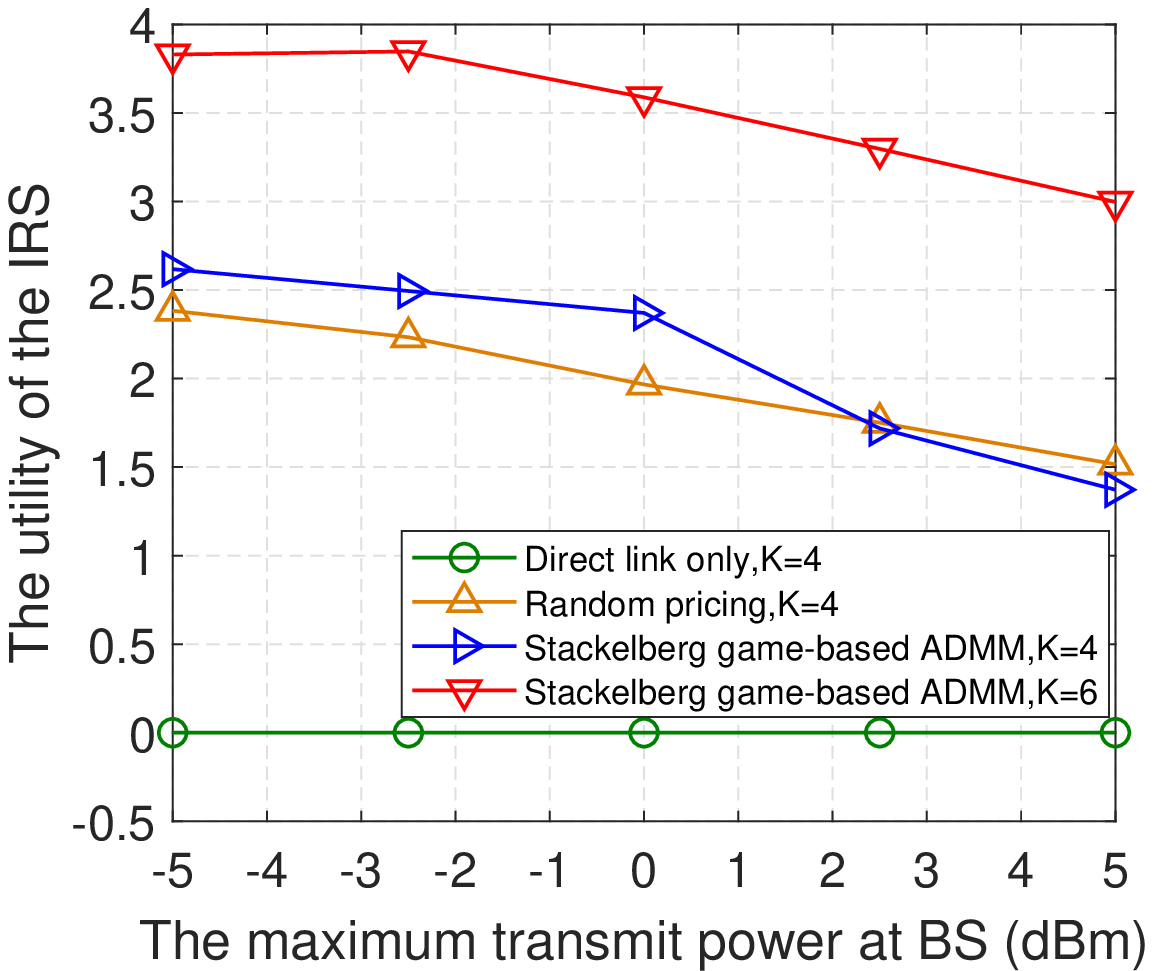}
		\caption{Impact of $p^{\max}$ on $V$.}
		\label{fig:2}
	\end{minipage}
\end{figure}

\begin{figure}[!t]
	\centering
	\begin{minipage}[t]{0.45\linewidth}
		\includegraphics[width=1.1\linewidth]{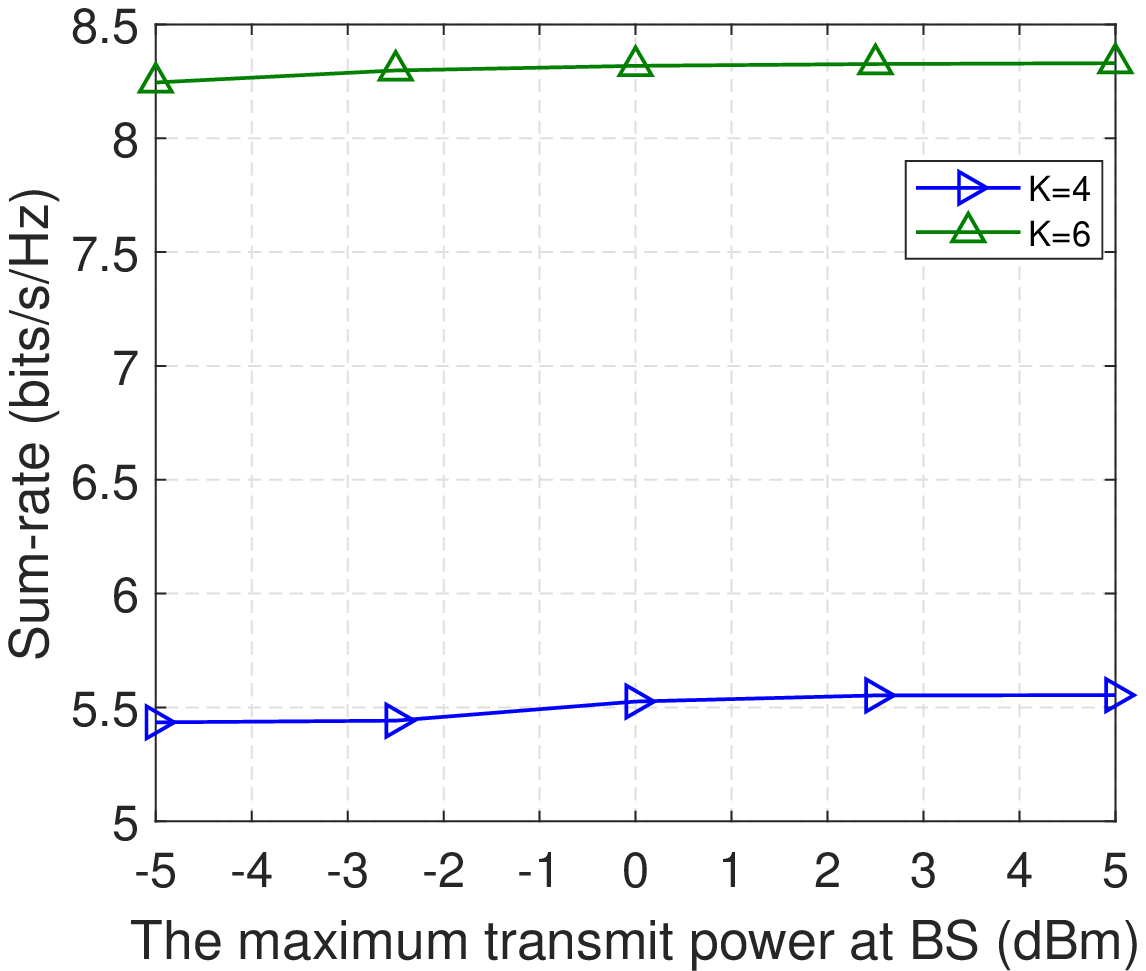}
		\caption{Sum rate  vs. $P^{\max}.$}
		\label{fig:3}
		\end{minipage}
	\centering
	\begin{minipage}[t]{0.45\linewidth}
	\includegraphics[width=1.1\linewidth]{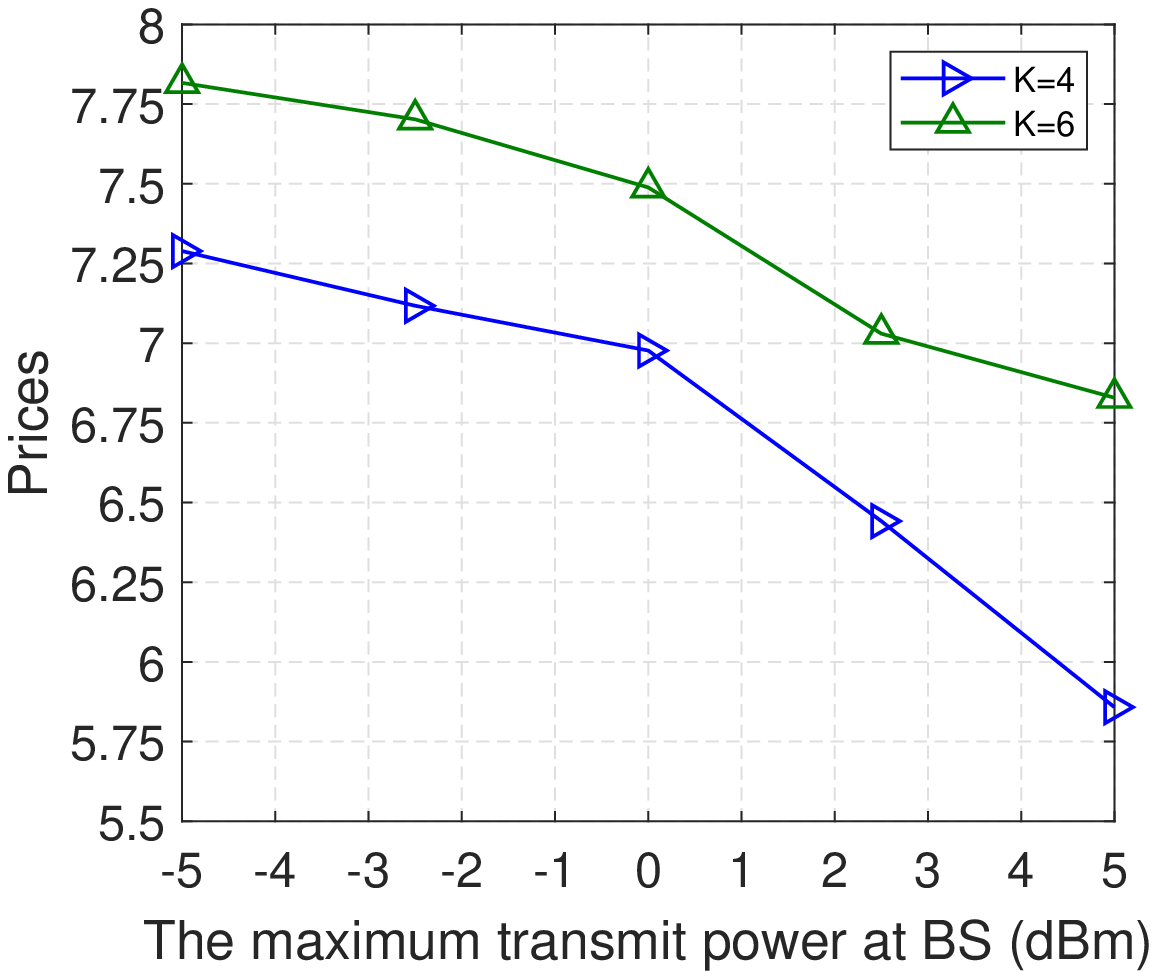}
	\caption{ Prices vs. $p^{\max}.$ }
	\label{fig:4}
	\end{minipage}
\end{figure}

The performance of the Stackelberg game-based ADMM scheme is evaluated against two existing benchmark schemes, i.e., random pricing scheme and direct link only scheme.
In the random pricing scheme, the IRS randomly determines its strategies, without considering the existence of the BS.
The direct link only scheme  means no IRS to aid, i.e., no module is activated at the IRS.
Figures \ref{fig:1} and \ref{fig:2} show the effect of the maximum transmit power $p^{\max}$ on the utility of the BS and the IRS, respectively, when the number of modules is $8.$
Correspondingly,  Figs. \ref{fig:3} and \ref{fig:4} depict  the sum rate of all users and the service prices versus the maximum transmit power at BS, respectively.
For the BS and IRS, the Stackelberg game-based ADMM scheme achieves the highest utility value compared with random pricing and direct link schemes, which indicates that the proposed pricing-based Stackelberg game scheme performs best in resource allocation for IRS-aided communications.
From the results, we observe that the utility values of the BS increases as $p^{\max}$ grows from $-5\text{~dBm}$ to $5\text{~dBm}.$
Meanwhile,  the utility value of the IRS achieved by the Stackelberg game-based ADMM scheme first decreases slowly until the maximum transmit power increases to $0\text{~dBm}$ and then decreases rapidly by increasing the value of $p^{\max}$.
This is because that the cost of power consumption is not considered in the utility of the BS, and thereby,  the BS will tend to select a small number of active modules when the transmit power is sufficient.
Meanwhile, for $p^{\max}>0\text{~dBm}$, the IRS needs to incentive the BS to select reflection resource through lower price, which can be observed from Fig. \ref{fig:4}.

\section{Conclusion}
The adoption of an IRS for downlink multi-user communication from a multi-antenna BS was investigated in this paper.
Specifically,  we developed a Stackelbeg game approach to analyze the interaction between the BS and the IRS operator considering that the IRS operator may be selfish or has its own objective.
Different from the existing studies on IRS that merely focused on tuning the reflection coefficient of all the reflecting elements, we considered the  reflection resource allocation, which can be realized via active module selection under the proposed modular IRS architecture that  all the modules are controlled by independent controllers.
The Stackelberg game-based ADMM was proposed to solve either the transmit beamforming at the BS or the passive beamorming of the activated modules.
Numerical examples were presented to verify the proposed studies.
It was shown that the proposed scheme is effective in the utilities of both the BS and IRS.
\section*{Acknowledgement}
{This work was supported in part by  the National Science Foundation of China under Grant number 61671131, also supported by the National Research Foundation (NRF), Singapore, under Singapore Energy Market Authority (EMA), Energy Resilience, NRF2017EWT-EP003-041, Singapore NRF2015-NRF-ISF001-2277, Singapore NRF National Satellite of Excellence, Design Science and Technology for Secure Critical Infrastructure NSoE DeST-SCI2019-0007, A*STAR-NTU-SUTD Joint Research Grant on Artificial Intelligence for the Future of Manufacturing RGANS1906, Wallenberg AI, Autonomous Systems and Software Program and Nanyang Technological University (WASP/NTU) under grant M4082187 (4080), Singapore Ministry of Education (MOE) Tier 1 (RG16/20), and NTU-WeBank JRI (NWJ-2020-004), Alibaba Group through Alibaba Innovative Research (AIR) Program, Alibaba-NTU Singapore Joint Research Institute (JRI),
	Nanyang Technological University (NTU) Startup Grant, Alibaba-NTU Singapore Joint Research Institute (JRI),
	Singapore Ministry of Education Academic Research Fund Tier 1 RG128/18, Tier 1 RG115/19, Tier 1 RT07/19, Tier 1 RT01/19, and Tier 2 MOE2019-T2-1-176,
	NTU-WASP Joint Project,
	Singapore National Research Foundation (NRF) under its Strategic Capability Research Centres Funding Initiative: Strategic Centre for Research in Privacy-Preserving Technologies \& Systems (SCRIPTS), Energy Research Institute @NTU (ERIAN),
	Singapore NRF National Satellite of Excellence,
	Design Science and Technology for Secure Critical Infrastructure NSoE DeST-SCI2019-0012,
	AI Singapore (AISG) 100 Experiments (100E) programme,
	NTU Project for Large Vertical Take-Off \& Landing (VTOL) Research Platform.}


\end{document}